\documentclass{PoS}

\usepackage{braket}
\usepackage[utf8]{inputenc}   % std linux encoding
\usepackage[sans]{dsfont}
\usepackage{eurosym}
\usepackage{array}
\usepackage{tabularx}
\usepackage{booktabs}
\usepackage{bm}
\usepackage{amsmath}
\usepackage{amssymb}
\usepackage{amsmath, amsfonts, epsfig, xspace}
\usepackage{listings}
\usepackage{algorithmic}
\usepackage{verbatim}
\graphicspath{{figures/}}
\usepackage{grffile}
\usepackage{indentfirst}

\usepackage{subfig}
\usepackage{float}
\usepackage{url}
\usepackage{listings}
\usepackage{algorithm}
\usepackage{algorithmic}
\usepackage{grffile}

\usepackage{adjustbox}
\usepackage{array}
\usepackage{booktabs}
\usepackage{multirow}
\usepackage{ragged2e}

\usepackage{cleveref}
\crefname{section}{Section}{Sections}
\Crefname{section}{Section}{Sections}
\crefname{equation}{Eq.}{Eqs.}
\Crefname{equation}{Eq.}{Eqs.}
\crefname{figure}{Fig.}{Figs.}
\Crefname{figure}{Fig.}{Figs.}
\crefname{table}{Table}{Tables}
\Crefname{table}{Table}{Tables}

\newcommand{\Tr}{\mbox{\rm Tr}}

\newcommand\T{\rule{0pt}{2.6ex}}
\newcommand\B{\rule[-1.2ex]{0pt}{0pt}}

\newlength{\arrayrulewidthOriginal}

\title{Flux tubes at Finite Temperature}
\ShortTitle{\ShortTitle{Short Title for header}}

\author{\speaker{Pedro Bicudo}\thanks{Portuguese Lattice QCD Collaboration}\\
              CFTP, Departamento de F\'{\i}sica, Instituto Superior T\'ecnico,
Universidade de Lisboa,\newline Avenida Rovisco Pais 1, 1049-001 Lisbon, Portugal \\
        E-mail: \email{bicudo@tecnico.ulisboa.pt}}

\author{Nuno Cardoso$^\dagger$\\
 CFTP, Departamento de F\'{\i}sica, Instituto Superior T\'ecnico,
Universidade de Lisboa,\newline Avenida Rovisco Pais 1, 1049-001 Lisbon, Portugal \\
        E-mail: \email{nunocardoso@cftp.ist.utl.pt}}

\author{Marco Cardoso$^\dagger$\\
 CFTP, Departamento de F\'{\i}sica, Instituto Superior T\'ecnico,
Universidade de Lisboa,\newline Avenida Rovisco Pais 1, 1049-001 Lisbon, Portugal \\
        E-mail: \email{mjdcc@cftp.ist.utl.pt}}

\abstract{
We show the flux tubes produced by static quark-antiquark,
quark-quark and quark-gluon charges at finite temperature.
The sources are placed in the lattice with fundamental and adjoint Polyakov loops.
We compute the square densities of the chromomagnetic and chromoelectric fields above and below the phase transition.
Our results are gauge invariant and produced in pure gauge SU(3).
The codes are written in CUDA and the computations are performed with GPUs.
}

\FullConference{34th annual International Symposium on Lattice Field Theory\\
		24-30 July 2016\\
		University of Southampton, UK}

\begin{document}

%SSSSSSSSSSSSSSSSSSSSSSSSSSSSSSSSSSSSSSSSSSSSSSSSSSSSSSSSSSSSSSSSSSSSSSSSSSSS
%SSSSSSSSSSSSSSSSSSSSSSSSSSSSSSSSSSSSSSSSSSSSSSSSSSSSSSSSSSSSSSSSSSSSSSSSSSSS
%SSSSSSSSSSSSSSSSSSSSSSSSSSSSSSSSSSSSSSSSSSSSSSSSSSSSSSSSSSSSSSSSSSSSSSSSSSSS
%SSSSSSSSSSSSSSSSSSSSSSSSSSSSSSSSSSSSSSSSSSSSSSSSSSSSSSSSSSSSSSSSSSSSSSSSSSSS
%SSSSSSSSSSSSSSSSSSSSSSSSSSSSSSSSSSSSSSSSSSSSSSSSSSSSSSSSSSSSSSSSSSSSSSSSSSSS
%SSSSSSSSSSSSSSSSSSSSSSSSSSSSSSSSSSSSSSSSSSSSSSSSSSSSSSSSSSSSSSSSSSSSSSSSSSSS
%SSSSSSSSSSSSSSSSSSSSSSSSSSSSSSSSSSSSSSSSSSSSSSSSSSSSSSSSSSSSSSSSSSSSSSSSSSSS
\section{ Introduction}

%----------------------------------------------------------------------------------------
\begin{figure}[t!]
\centering
\includegraphics[width=0.9\columnwidth]{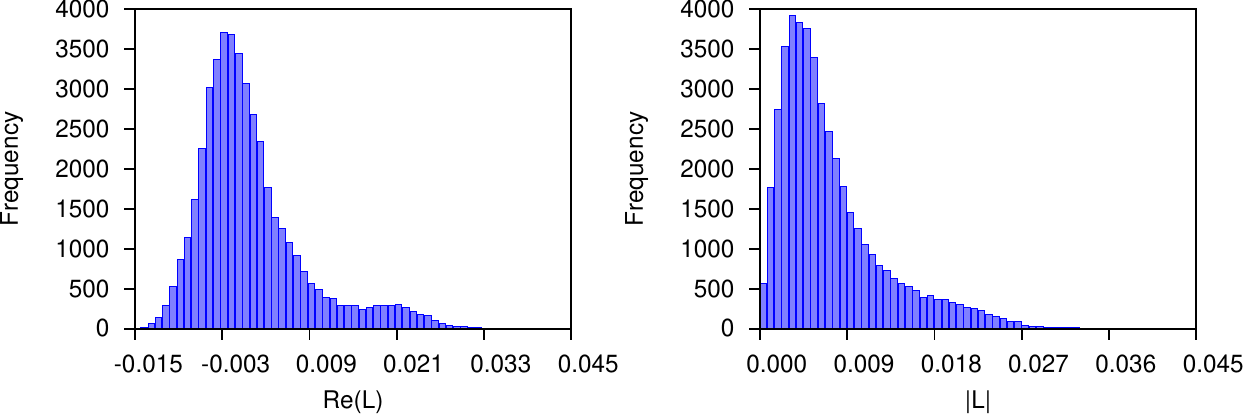}
\caption{Histogram of the Polyakov loop history for $\beta=6.055$.}
\label{histpolyloop}
\end{figure}

We study of the colour fields distributions inside the flux tubes formed by Polyakov loops in the static $Q\bar{Q}$, $QQ$ and $QG$ systems.
We address how the flux tube evolves with the distance between quarks and when the temperature increase beyond the deconfinement temperature.
%In this paper, we study the chromo fields of the $QQ$ and $q\bar{Q}$ systems below and above the phase transition. 
In section 2, we describe the lattice formulation. We briefly review the Polyakov loop for these systems and show how to compute the colour fields as well as the Lagrangian distribution. In section 3, the numerical results are shown. Finally, we conclude in section 4.

\begin{table}[b!]
\begin{center}
\begin{tabular}{|cccc|}
%\hline
\hline
\T\B $\beta$ & $T/T_c$	&	$a\sqrt{\sigma}$	&	\# config.	\\ \hline
\hline
\T\B 5.96 & 0.845	&	0.235023	&	5990	\\ \hline
\T\B 6.0534 & 0.986	&	0.201444	&	5990/5110* 	\\ \hline
\T\B 6.13931 & 1.127	&	0.176266	&	5990 	\\ \hline
\T\B 6.29225 & 1.408	&	0.141013	&	5990 	\\ \hline
\T\B 6.4249 & 1.690	&	0.117513	&	5990 	\\ \hline
%\hline
\end{tabular}
\end{center}
\caption{Lattice simulations for a $48^3\times 8$ volume.
We denote with an $^*$ the number of remaining configurations after we remove the configurations in the other phase.}
\label{tab:latticesimdata}
\end{table}

%SSSSSSSSSSSSSSSSSSSSSSSSSSSSSSSSSSSSSSSSSSSSSSSSSSSSSSSSSSSSSSSSSSSSSSSSSSSS
%SSSSSSSSSSSSSSSSSSSSSSSSSSSSSSSSSSSSSSSSSSSSSSSSSSSSSSSSSSSSSSSSSSSSSSSSSSSS
%SSSSSSSSSSSSSSSSSSSSSSSSSSSSSSSSSSSSSSSSSSSSSSSSSSSSSSSSSSSSSSSSSSSSSSSSSSSS
%SSSSSSSSSSSSSSSSSSSSSSSSSSSSSSSSSSSSSSSSSSSSSSSSSSSSSSSSSSSSSSSSSSSSSSSSSSSS
%SSSSSSSSSSSSSSSSSSSSSSSSSSSSSSSSSSSSSSSSSSSSSSSSSSSSSSSSSSSSSSSSSSSSSSSSSSSS
%SSSSSSSSSSSSSSSSSSSSSSSSSSSSSSSSSSSSSSSSSSSSSSSSSSSSSSSSSSSSSSSSSSSSSSSSSSSS
%SSSSSSSSSSSSSSSSSSSSSSSSSSSSSSSSSSSSSSSSSSSSSSSSSSSSSSSSSSSSSSSSSSSSSSSSSSSS
\section{ Computation of the chromo-fields in the flux tube}

%----------------------------------------------------------------------------------------
\begin{figure}[t!]
	\captionsetup[subfloat]{farskip=0.5pt,captionskip=0.5pt}
\begin{centering}

\hspace{-15pt}
\subfloat[$\beta=5.96$, $T=0.845T_c$.\label{fig:F00_5.96_ppdagger_}]{
\begin{centering}
\includegraphics[trim=0 30 10 0, clip,width=0.50\columnwidth]{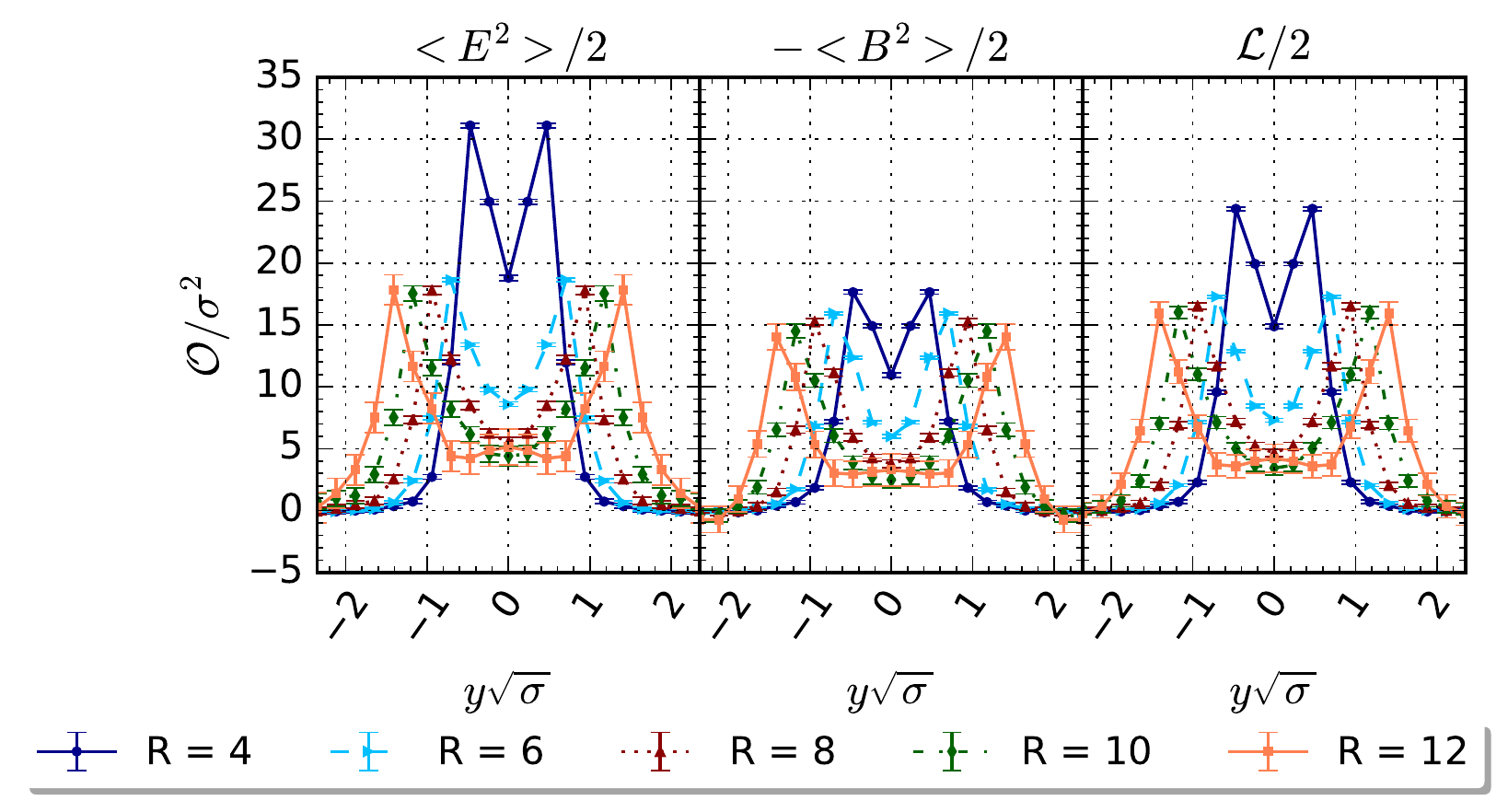}
\includegraphics[trim=0 30 10 0, clip,width=0.50\columnwidth]{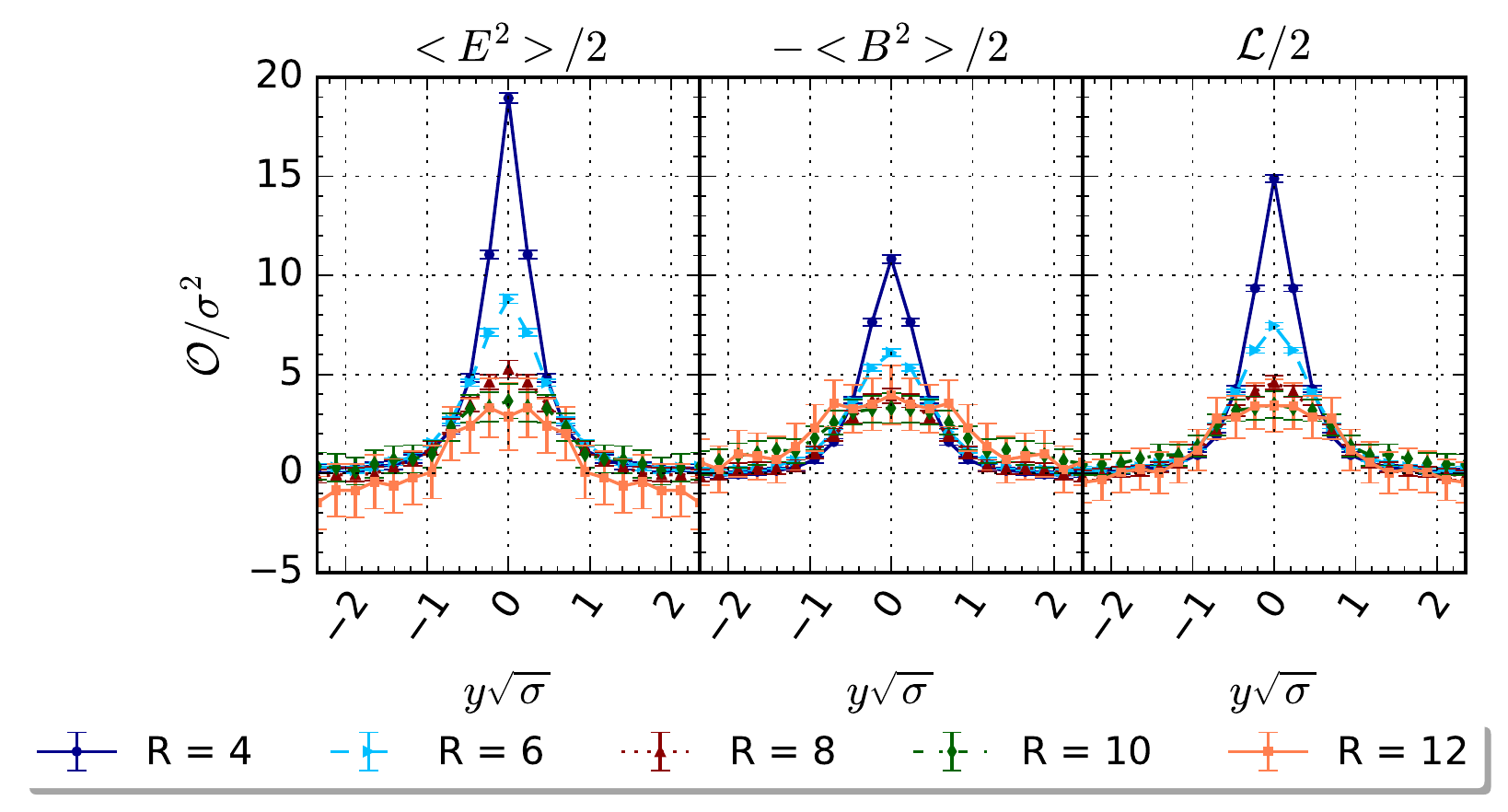}
\par\end{centering}}

\hspace{-15pt}
\subfloat[$\beta=6.0534$, $T=0.986T_c$, with contaminated configurations.\label{fig:F00_6.0534_ppdagger_}]{
\begin{centering}
\includegraphics[trim=0 30 10 0, clip,width=0.50\columnwidth]{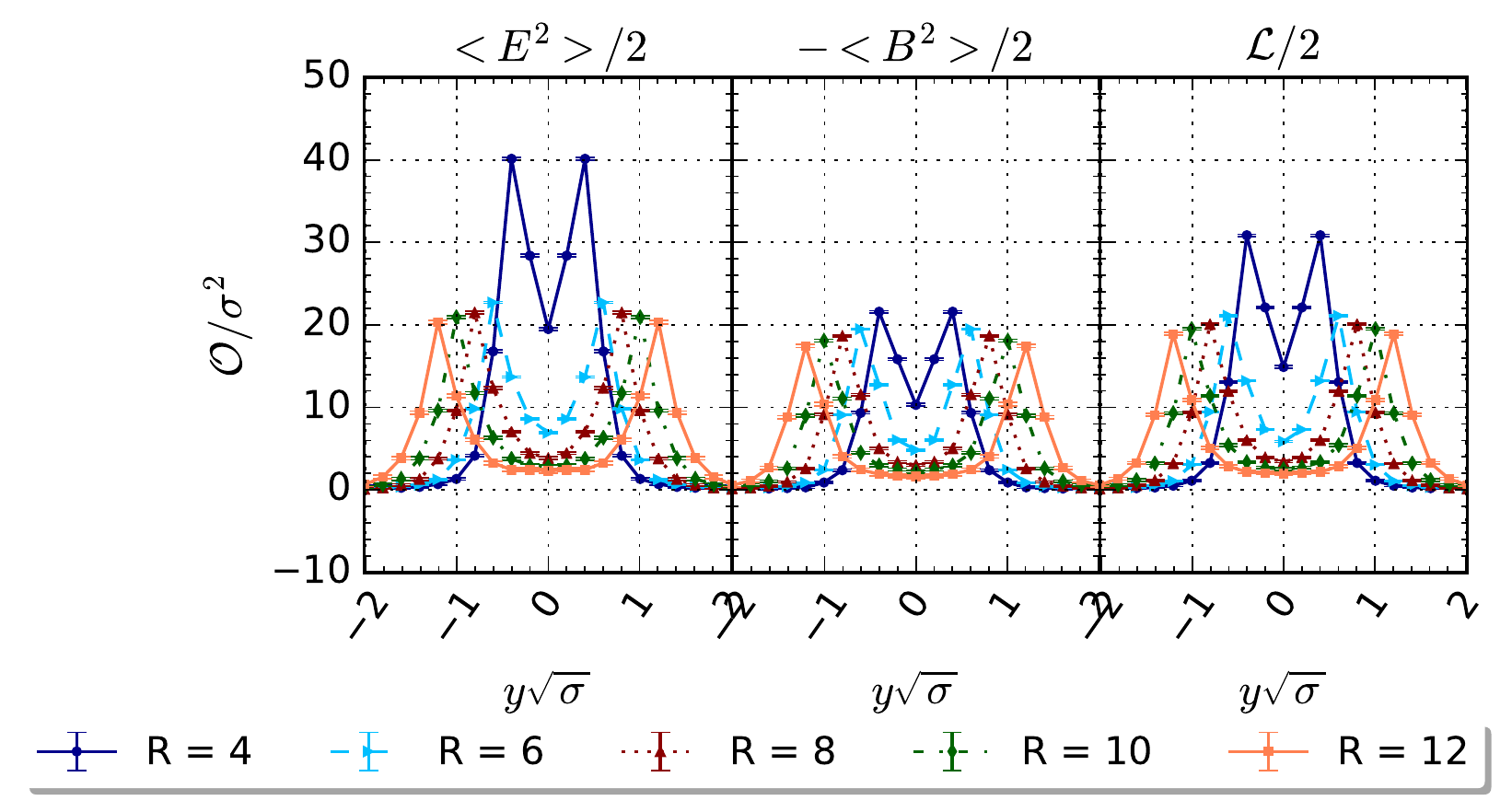}
\includegraphics[trim=0 30 10 0, clip,width=0.50\columnwidth]{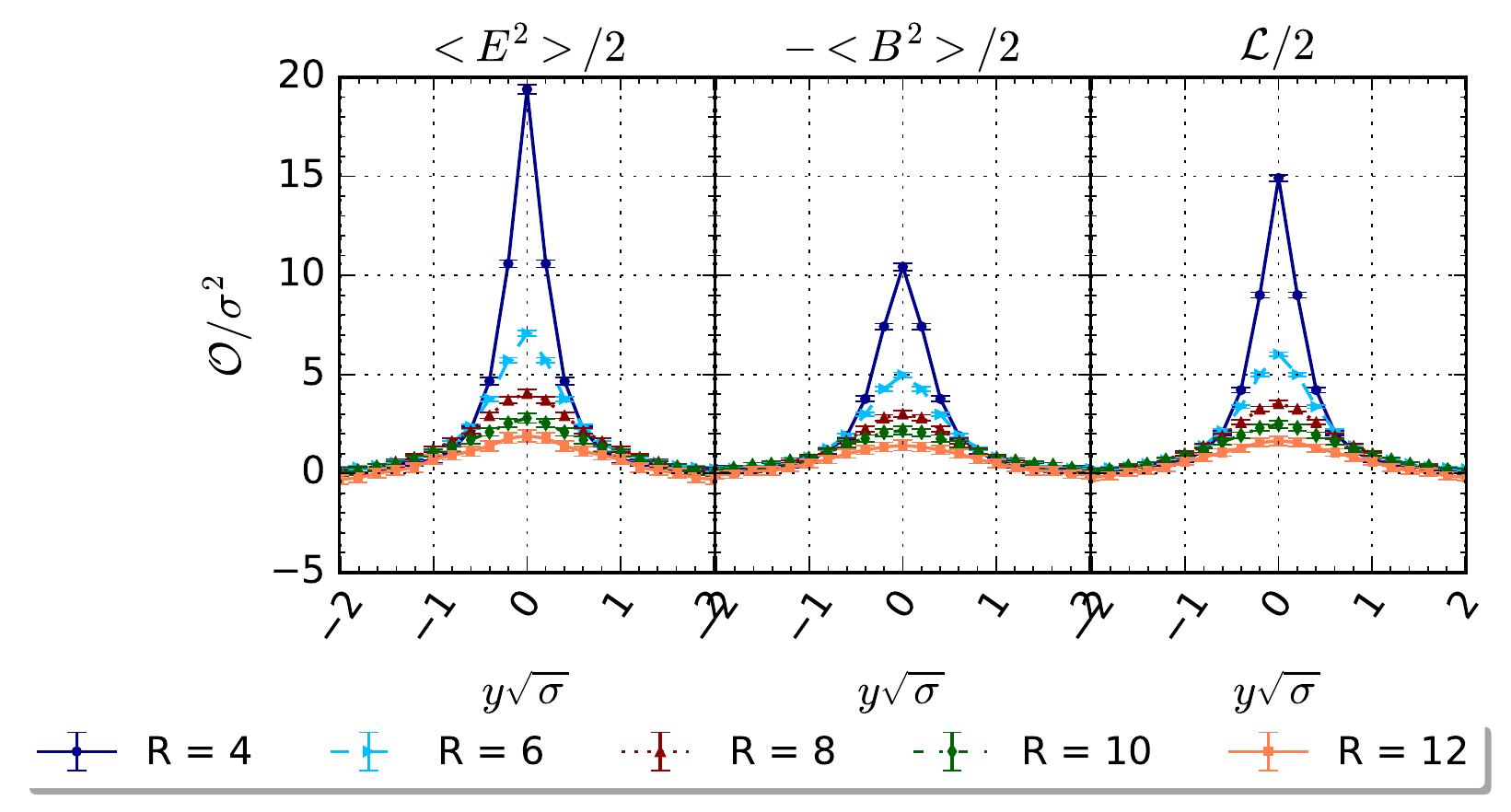}
\par\end{centering}}

\hspace{-15pt}
\subfloat[$\beta=6.0534$, $T=0.986T_c$, without contaminated configurations.\label{fig:F00_clean_6.0534_ppdagger_}]{
\begin{centering}
\includegraphics[trim=0 30 10 0, clip,width=0.50\columnwidth]{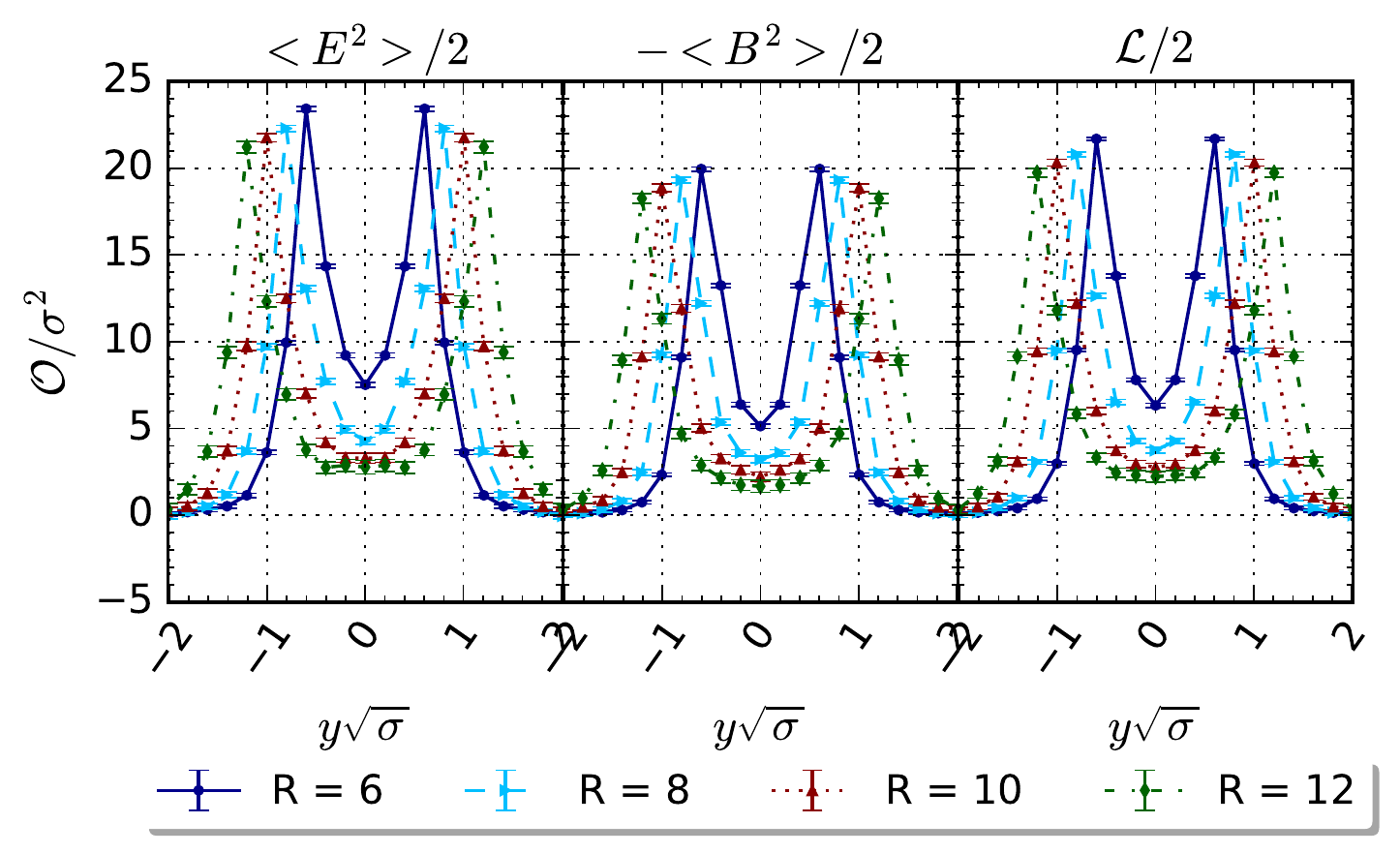}
\includegraphics[trim=0 30 10 0, clip,width=0.50\columnwidth]{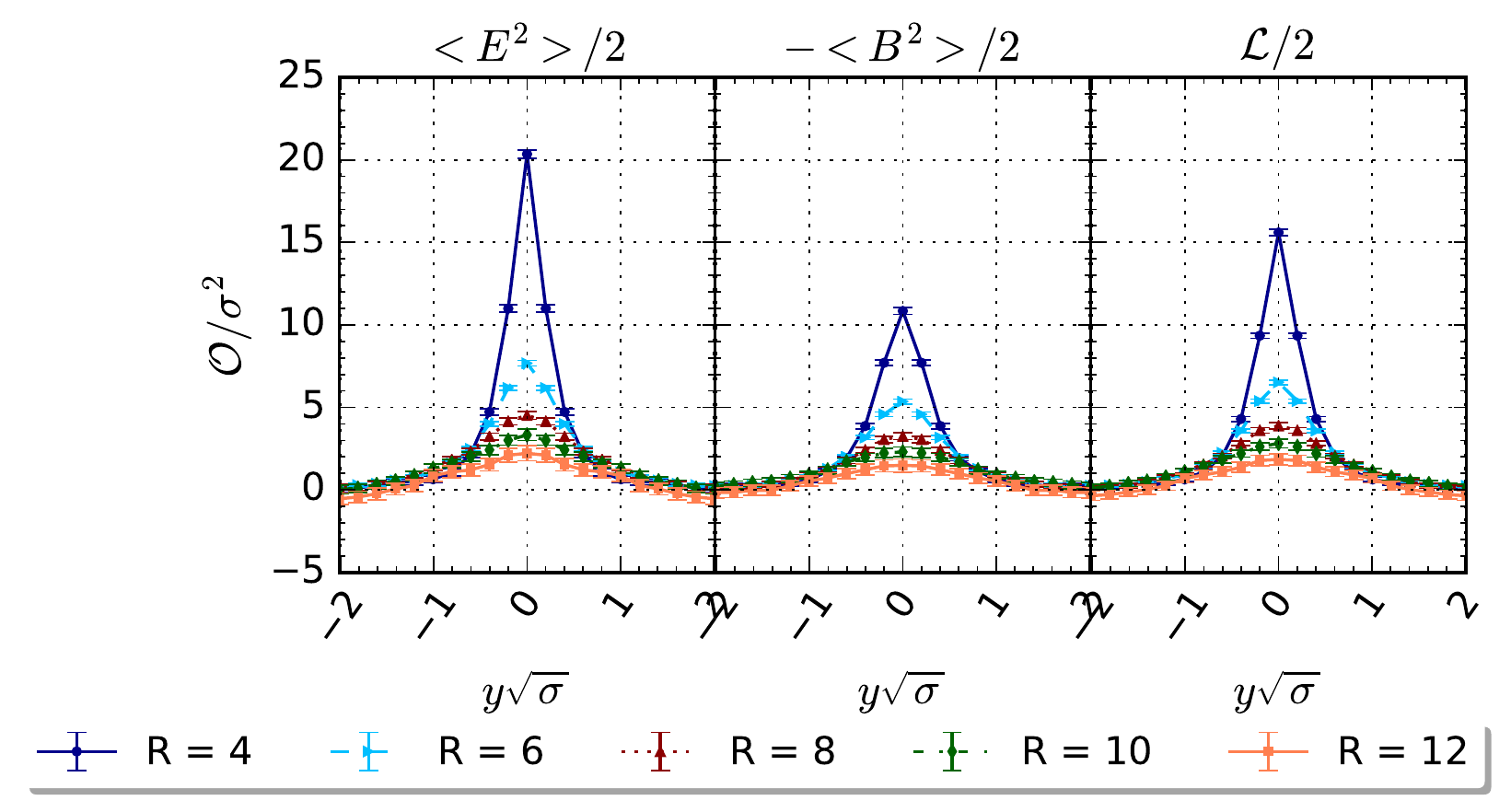}
\par\end{centering}}

\includegraphics[trim=5 0 10 260, clip,width=0.60\columnwidth]{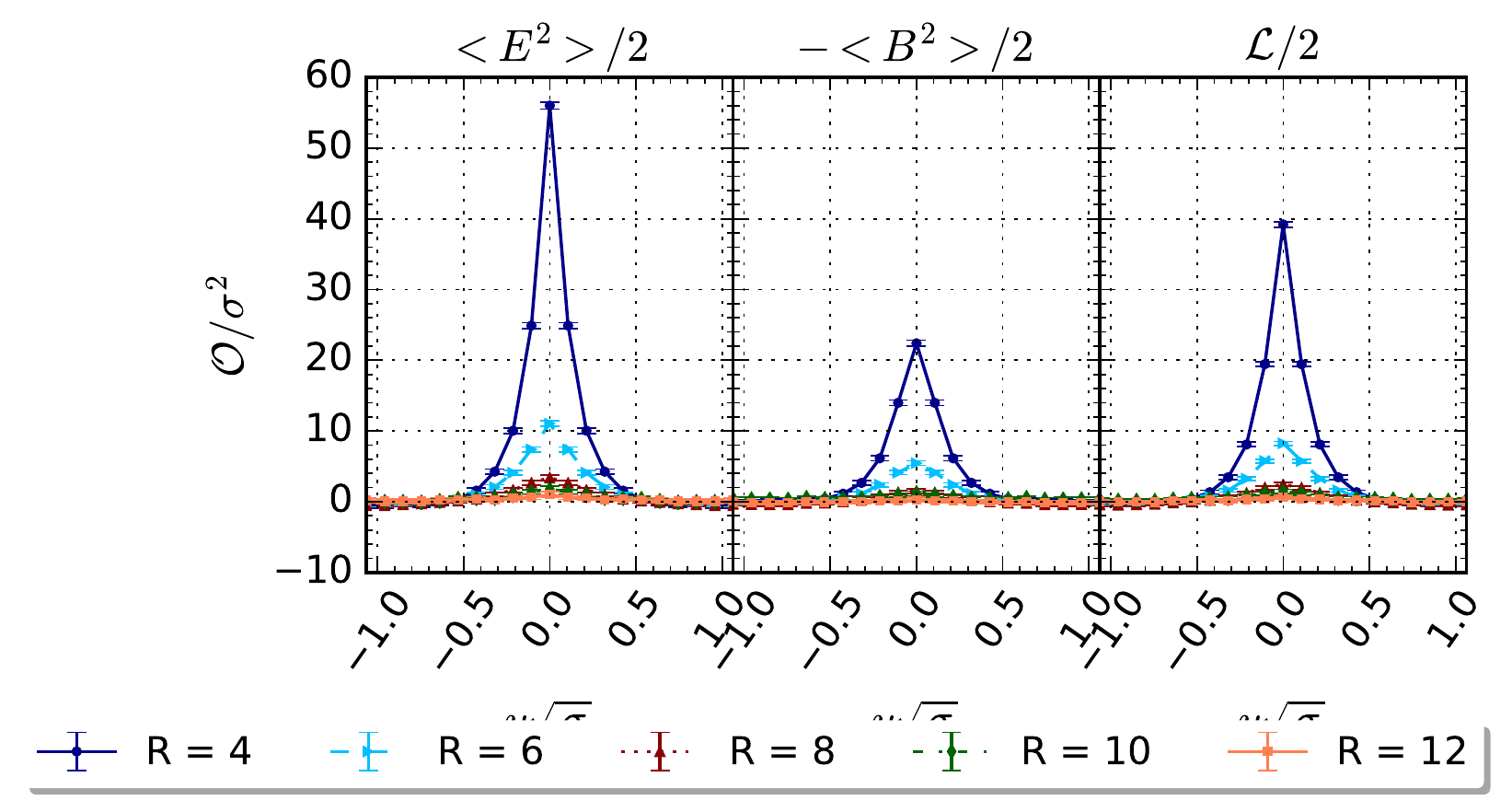}
\par\end{centering}
\caption{The results for the $Q\bar{Q}$ system at $T<T_c$. The results in the left column correspond to the fields along the sources (plane XY) and the right column to the results in the middle of the flux tube (plane XZ). $R$ is the distance between the sources in lattice units.}
\label{fig:shapeFluxTube_QQbar_Tless}
\end{figure}

The central observables that govern the event in the flux tube can be extracted from the correlation of a plaquette, $\square_{\mu\nu}$, with the Polyakov loops, $L$. To reduce the fluctuations of the $\mathcal{O}\,\square_{\mu\nu}(x)$, we measure the following quantity, \cite{PhysRevD.47.5104},
\begin{equation}
f_{\mu\nu}(r,x) = \frac{\beta}{a^4} \left[\frac{\Braket{\mathcal{O}\,\square_{\mu\nu}(x)}-\Braket{\mathcal{O}\,\square_{\mu\nu}(x_R)}}{\Braket{\mathcal{O}}}\right] \ ,
\label{eq:fmunucomp}
\end{equation}
where $x_R$ is the reference point placed far from the sources, and 
\begin{eqnarray}
&\mathcal{O}=L(0)\,L^\dagger(r)\quad\quad& \text{for the } Q\bar{Q} \text{ system} \ ,
\nonumber
\\
\nonumber
&\mathcal{O}=L(0)\,L(r)\quad\quad& \text{for the } QQ \text{ system} \ , 
\\
&\mathcal{O}=\left(L(0)L^\dagger(0)-1\right)\,L(r)\quad\quad& \text{for the } Qg \text{ system} \ .
\label{eq:polyakov_loops}
\end{eqnarray}
Moreover $x$ denotes the distance of the plaquette from the line connecting sources, $r$ is the separation between the sources, $L(r)=\frac{1}{3}\Tr\Pi_{t=1}^{N_t}U_4(r,t)$
and $N_t$ is the number of time slices of the lattice. We also use the periodicity in the time direction for the Polyakov loops, 
$\square_{\mu\nu}(x) = \frac{1}{N_t} \sum_{t=1}^{N_t} \square_{\mu\nu}(x,t)$, to average the plaquette over the time direction.
% $\square_{\mu\nu}(x) = \frac{1}{3} \Tr\left[U_\mu(x)U_\nu(x+\mu)U^\dagger_\mu(x+\nu)U^\dagger_\nu(x)\right]$.
%The periodicity in the time direction also allows averaging over the time direction,

%----------------------------------------------------------------------------------------
\begin{figure}[t!]
	\captionsetup[subfloat]{farskip=0.5pt,captionskip=0.5pt}
\begin{centering}

\hspace{-15pt}
\subfloat[$\beta=6.13931$, $T=1.127T_c$.\label{fig:F00_6.13931_ppdagger_}]{
\begin{centering}
\includegraphics[trim=0 25 10 0, clip,width=0.50\columnwidth]{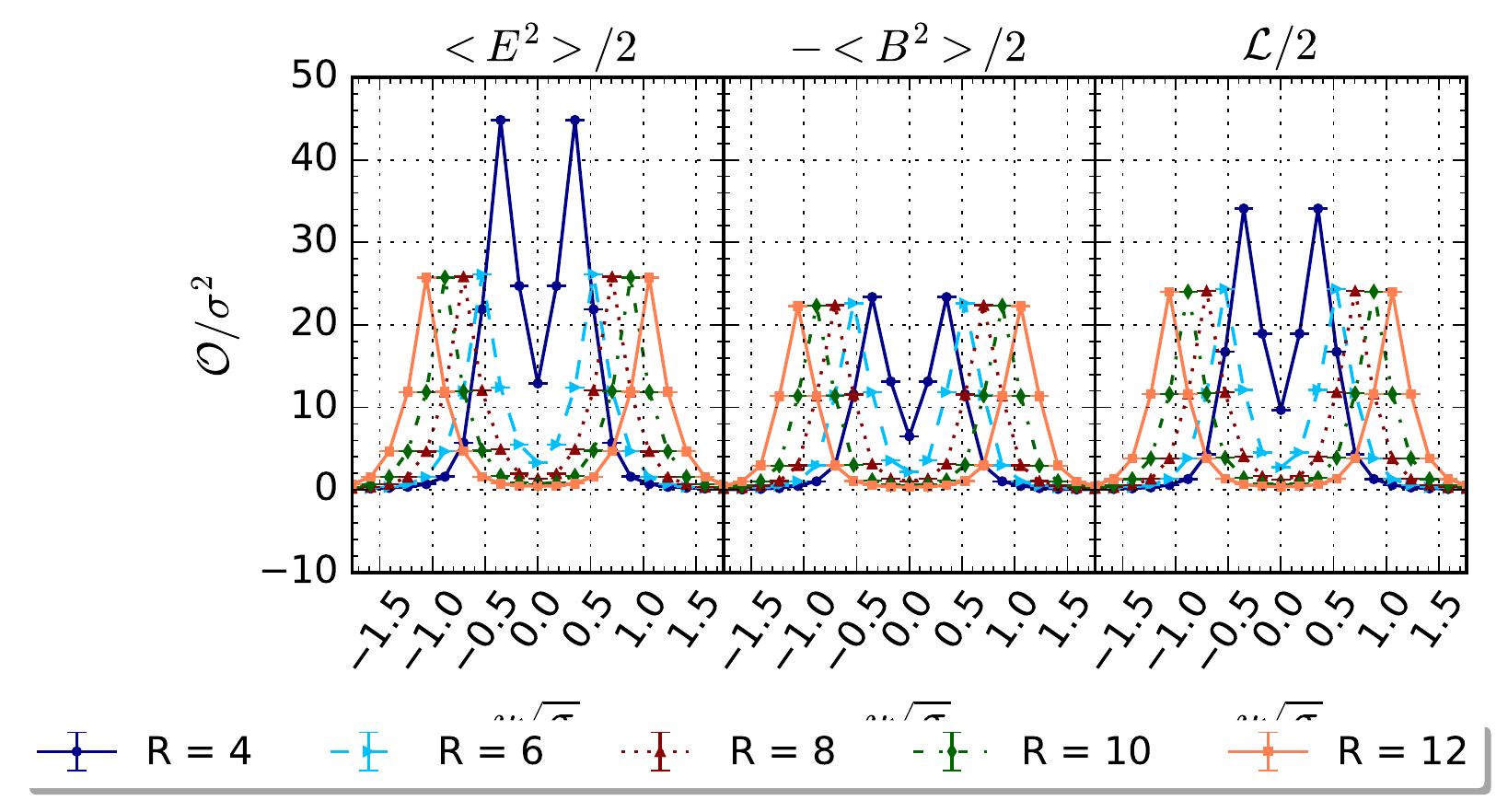}
\includegraphics[trim=0 30 10 0, clip,width=0.50\columnwidth]{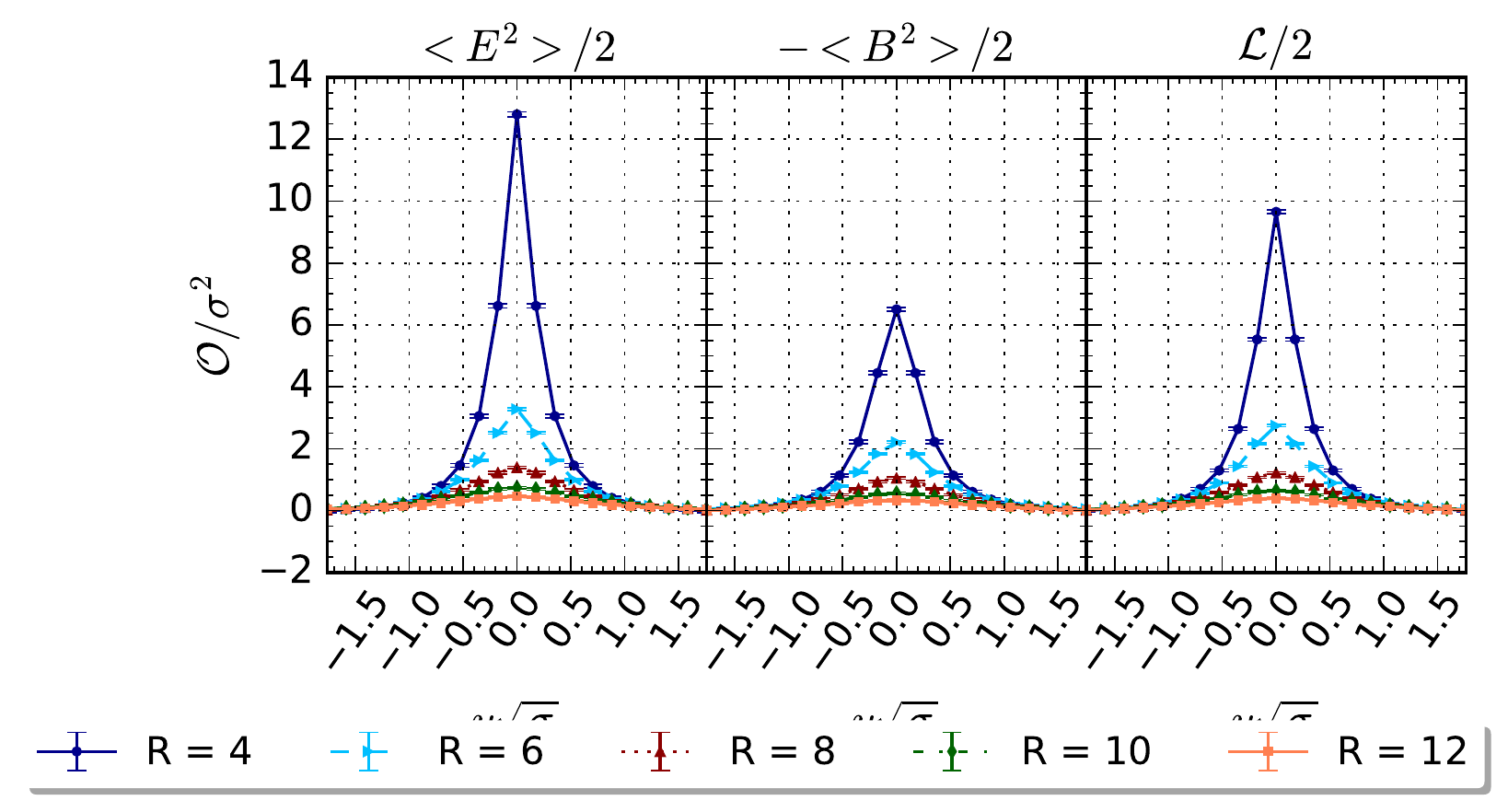}
\par\end{centering}}

\hspace{-15pt}
\subfloat[$\beta=6.29225$, $T=1.408T_c$.\label{fig:F00_6.292255_ppdagger_}]{
\begin{centering}
\includegraphics[trim=0 30 10 0, clip,width=0.50\columnwidth]{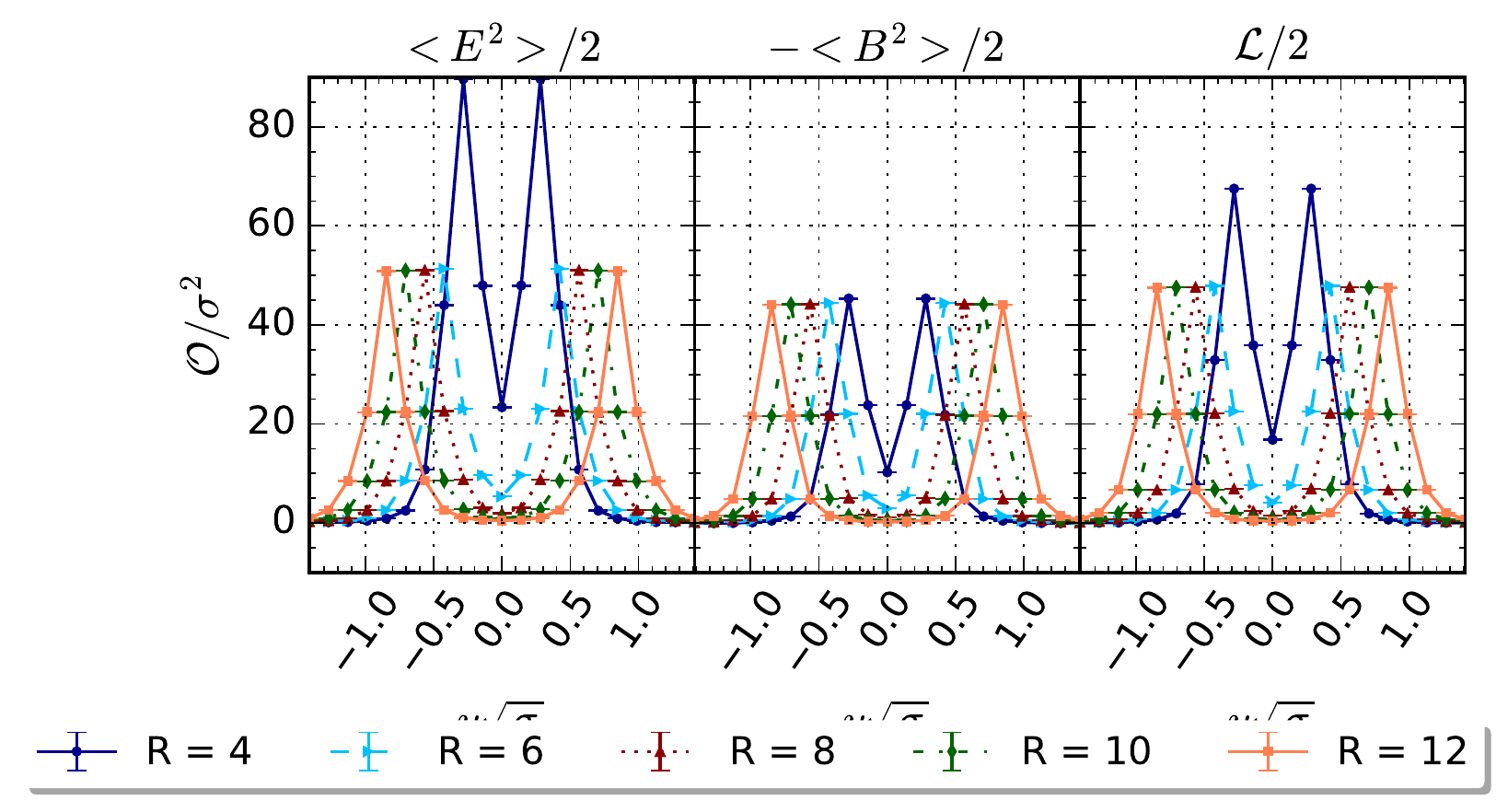}
\includegraphics[trim=0 30 10 0, clip,width=0.50\columnwidth]{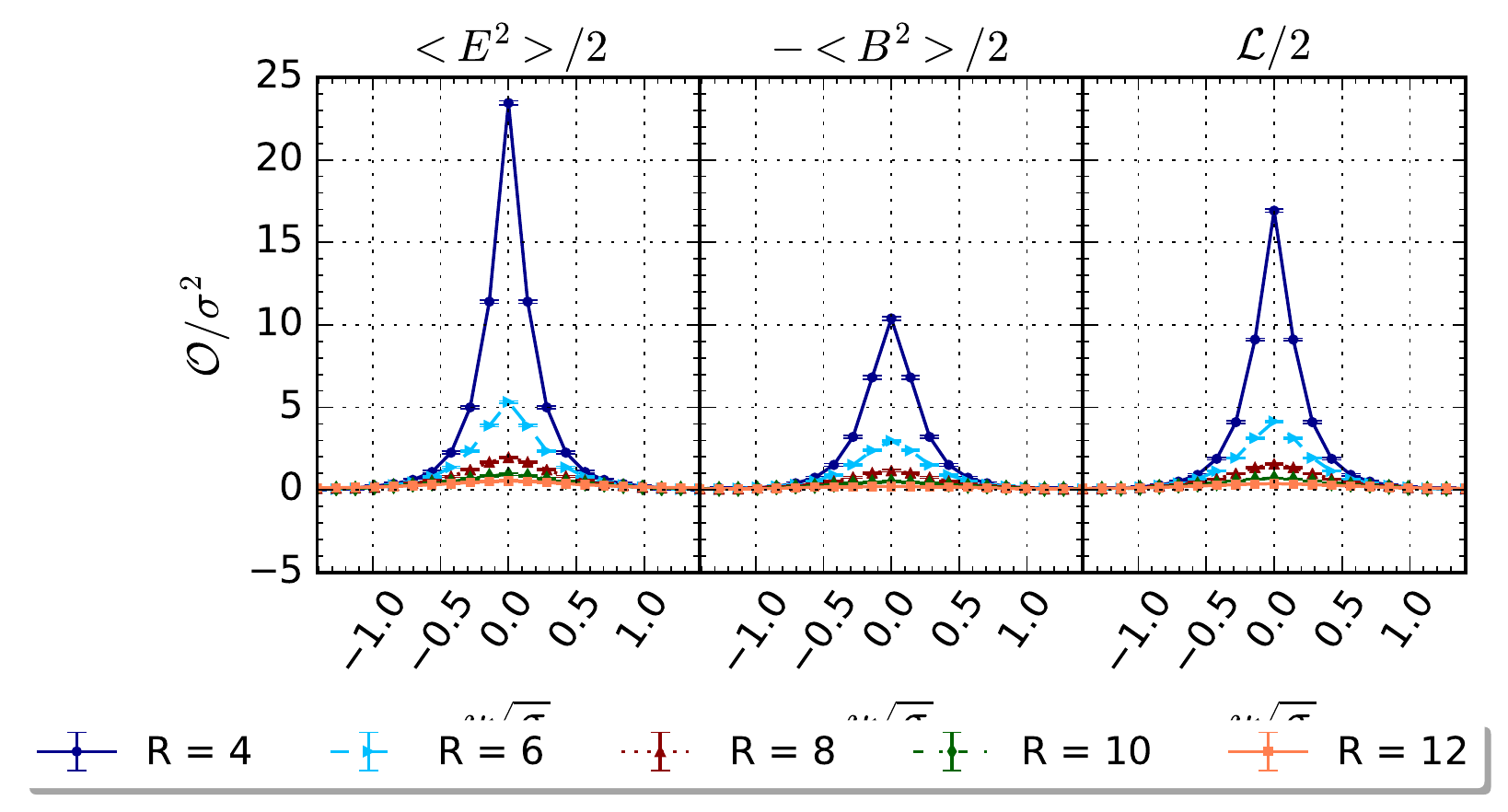}
\par\end{centering}}

\hspace{-15pt}
\subfloat[$\beta=6.4249$, $T=1.690T_c$.\label{fig:F00_6.4249_ppdagger_}]{
\begin{centering}
\includegraphics[trim=0 30 10 0, clip,width=0.50\columnwidth]{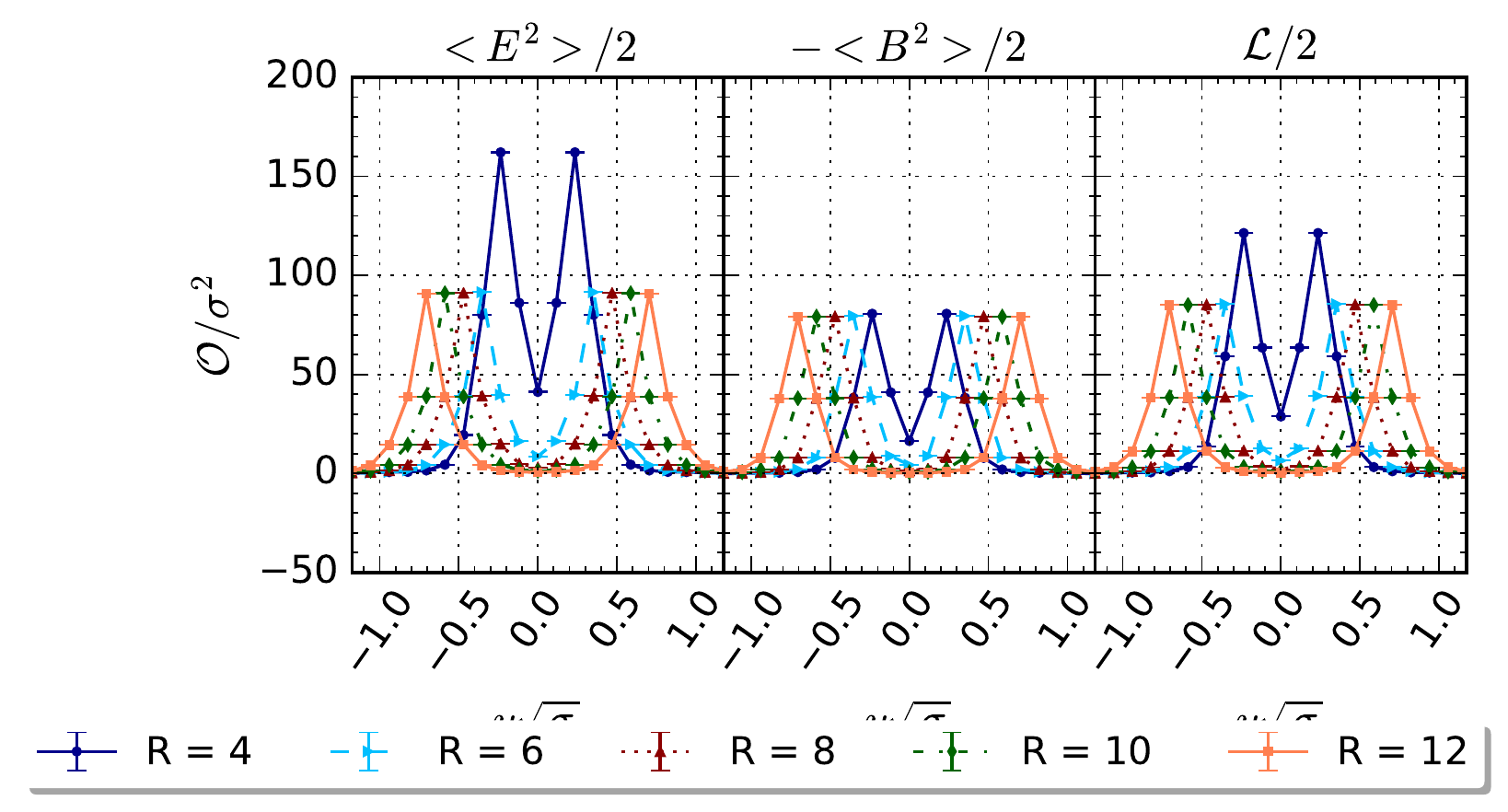}
\includegraphics[trim=0 30 10 0, clip,width=0.50\columnwidth]{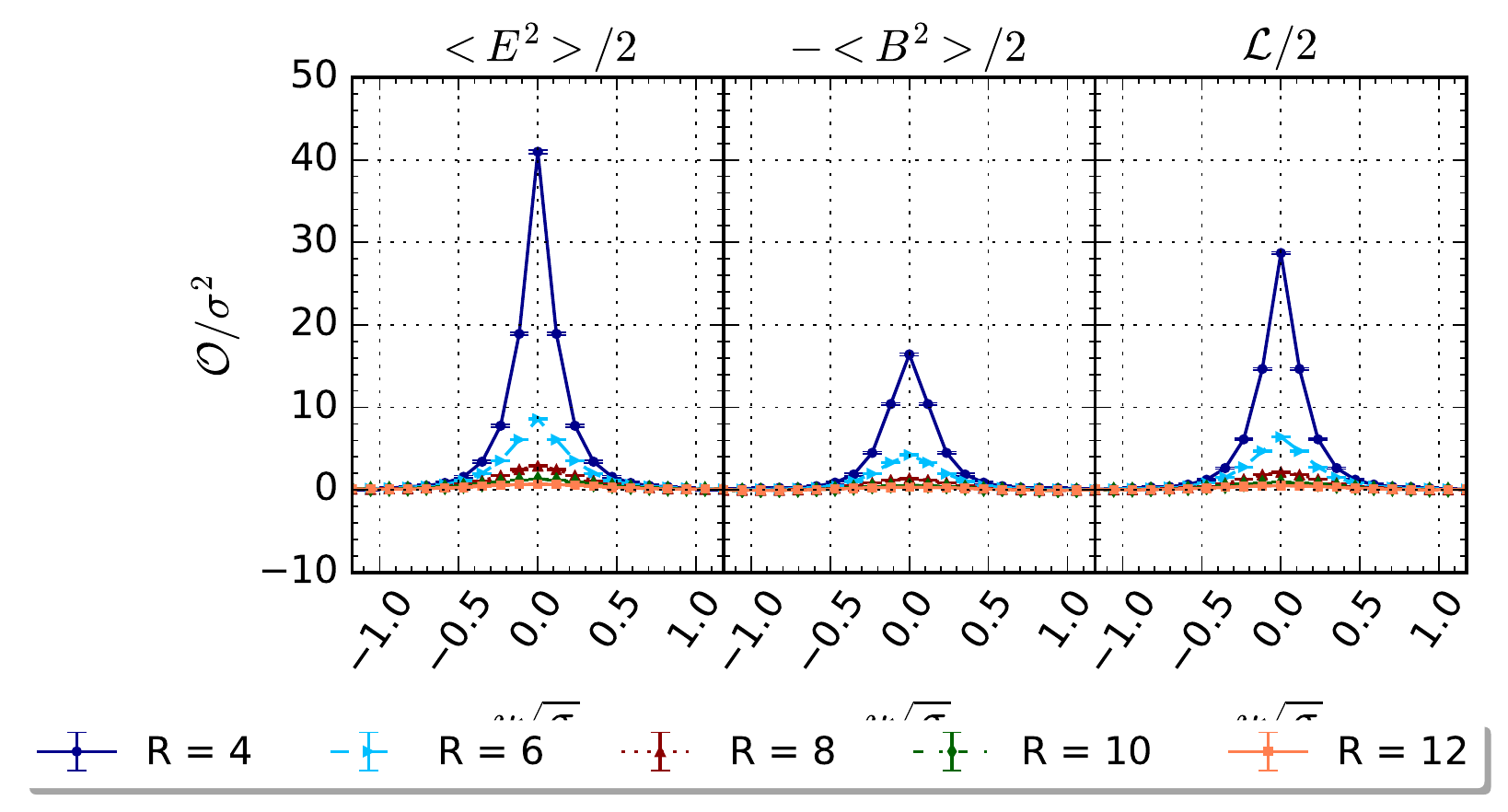}
\par\end{centering}}

\includegraphics[trim=5 0 10 260, clip,width=0.60\columnwidth]{F11_6.5_XZ_ppdagger_}
\par\end{centering}
\caption{The results for the $Q\bar{Q}$ system atr $T>T_c$. The results in the left column correspond to the fields along the sources (plane XY) and the right column to the results in the middle of the flux tube (plane XZ). $R$ is the distance between the sources in lattice units.}
\label{fig:shapeFluxTube_QQbar_Tmore}
\end{figure}

Therefore, using the plaquette orientation $(\mu,\nu)=(2,3), (1,3), (1,2),$ $(1,4),(2,4),(3,4)$, we can relate the six components in \cref{eq:fmunucomp} to the components of the chromoelectric and chromomagnetic fields,
\begin{equation}
f_{\mu\nu}\rightarrow\frac{1}{2}\left(-\Braket{B_x^2},-\Braket{B_y^2},-\Braket{B_z^2},\Braket{E_x^2},\Braket{E_y^2},\Braket{E_z^2}\right) \ ,
\end{equation}
and also calculate the total action (Lagrangian) density, $\Braket{\mathcal{L}}=\frac12\left(\Braket{E^2}-\Braket{B^2}\right)$.

In order to improve the signal over noise ratio in the $Q\bar{Q}$ and $QQ$ systems, we use the multihit technique,  \cite{Brower:1981vt, Parisi:1983hm}, replacing each temporal link by it's thermal average,
%\begin{equation}
%	U_4\rightarrow \bar{U}_4=\frac{\int dU_4 U_4 \,e^{\beta\Tr \left[U_4 F^\dagger\right]}}{\int dU_4 \,e^{\beta\Tr\left[ U_4 F^\dagger\right]}}
%\end{equation}
and the extended multihit technique, \cite{Cardoso:2013lla}.

%----------------------------------------------------------------------------------------
\begin{figure}[t!]
	\captionsetup[subfloat]{farskip=0.5pt,captionskip=0.5pt}
\begin{centering}
%\subfloat[$\beta=5.96$, $T=0.845T_c$.\label{fig:F00_5.96_pp_}]{
%\begin{centering}
%\includegraphics[trim=0 30 10 0, clip,width=0.33\columnwidth]{F11_5.96_XY_pp_}
%\includegraphics[trim=0 30 10 0, clip,width=0.33\columnwidth]{F11_5.96_XZ_pp_}
%\par\end{centering}}
%
%\subfloat[$\beta=6.0534$, $T=0.986T_c$, with contaminated configurations.\label{fig:F00_6.0534_pp_}]{
%\begin{centering}
%\includegraphics[trim=0 30 10 0, clip,width=0.33\columnwidth]{F11_6.0534_XY_pp_}
%\includegraphics[trim=0 30 10 0, clip,width=0.33\columnwidth]{F11_6.0534_XZ_pp_}
%\par\end{centering}}
%
%\subfloat[$\beta=6.0534$, $T=0.986T_c$, without contaminated configurations.\label{fig:F00_clean_6.0534_pp_}]{
%\begin{centering}
%\includegraphics[trim=0 30 10 0, clip,width=0.33\columnwidth]{F00_clean_6.055_XY_pp_}
%\includegraphics[trim=0 30 10 0, clip,width=0.33\columnwidth]{F00_clean_6.055_XZ_pp_}
%\par\end{centering}}
%
	\hspace{-15pt}
\subfloat[$\beta=6.13931$, $T=1.127T_c$.\label{fig:F00_6.13931_pp_}]{
\begin{centering}
\includegraphics[trim=0 30 10 0, clip,width=0.50\columnwidth]{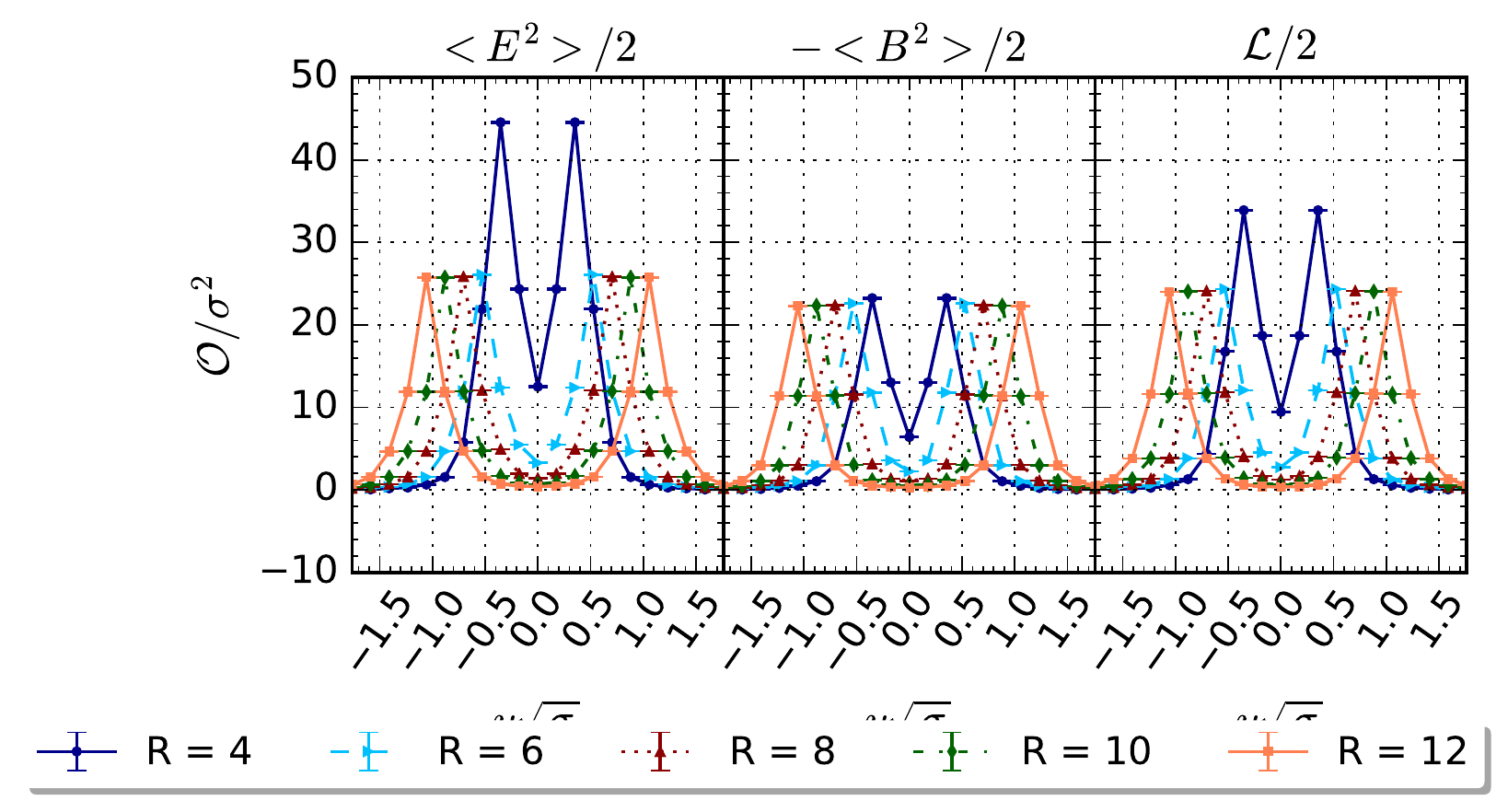}
\includegraphics[trim=0 30 10 0, clip,width=0.50\columnwidth]{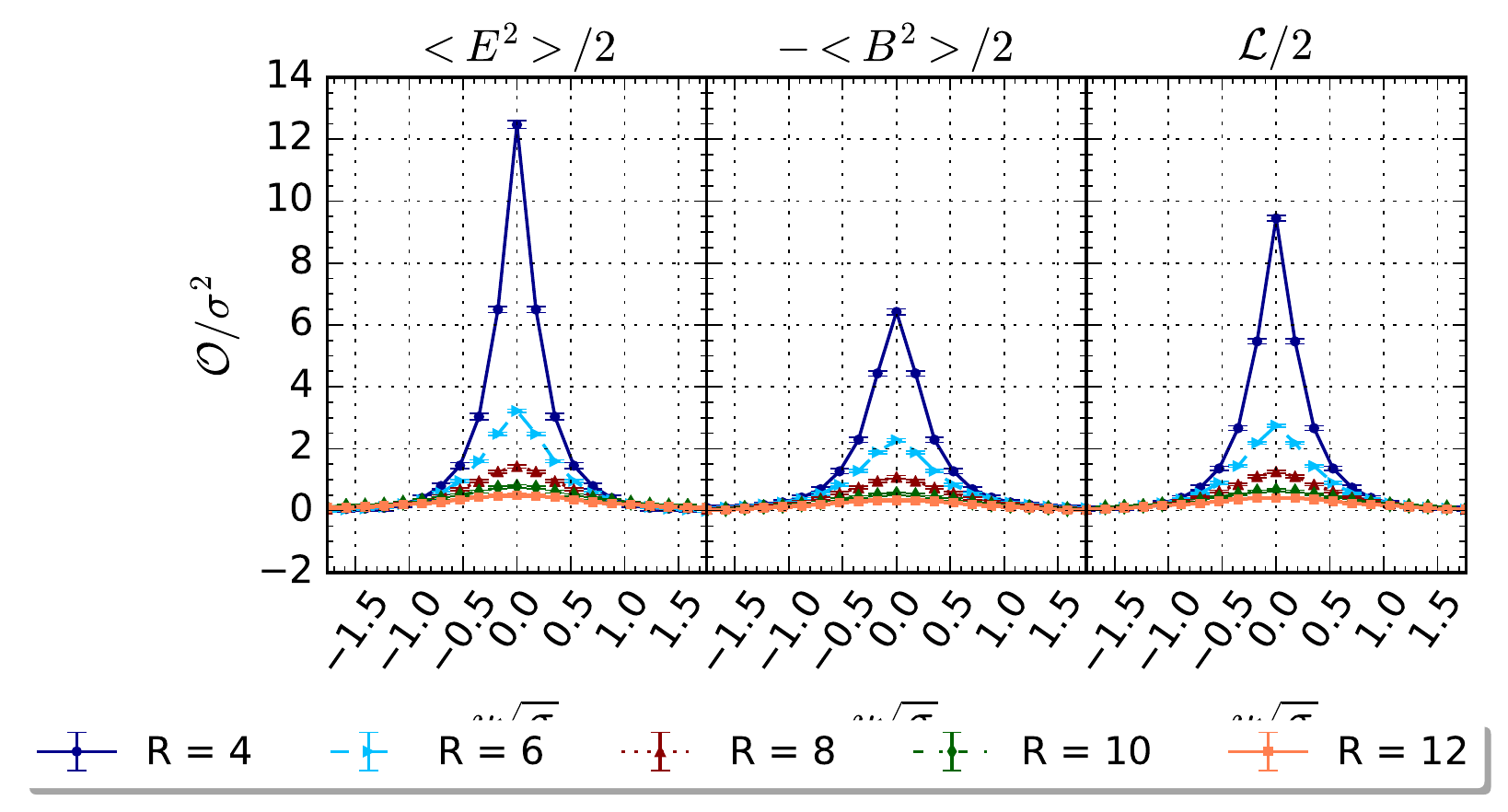}
\par\end{centering}}

	\hspace{-15pt}
\subfloat[$\beta=6.29225$, $T=1.408T_c$.\label{fig:F00_6.292255_pp_}]{
\begin{centering}
\includegraphics[trim=0 30 10 0, clip,width=0.50\columnwidth]{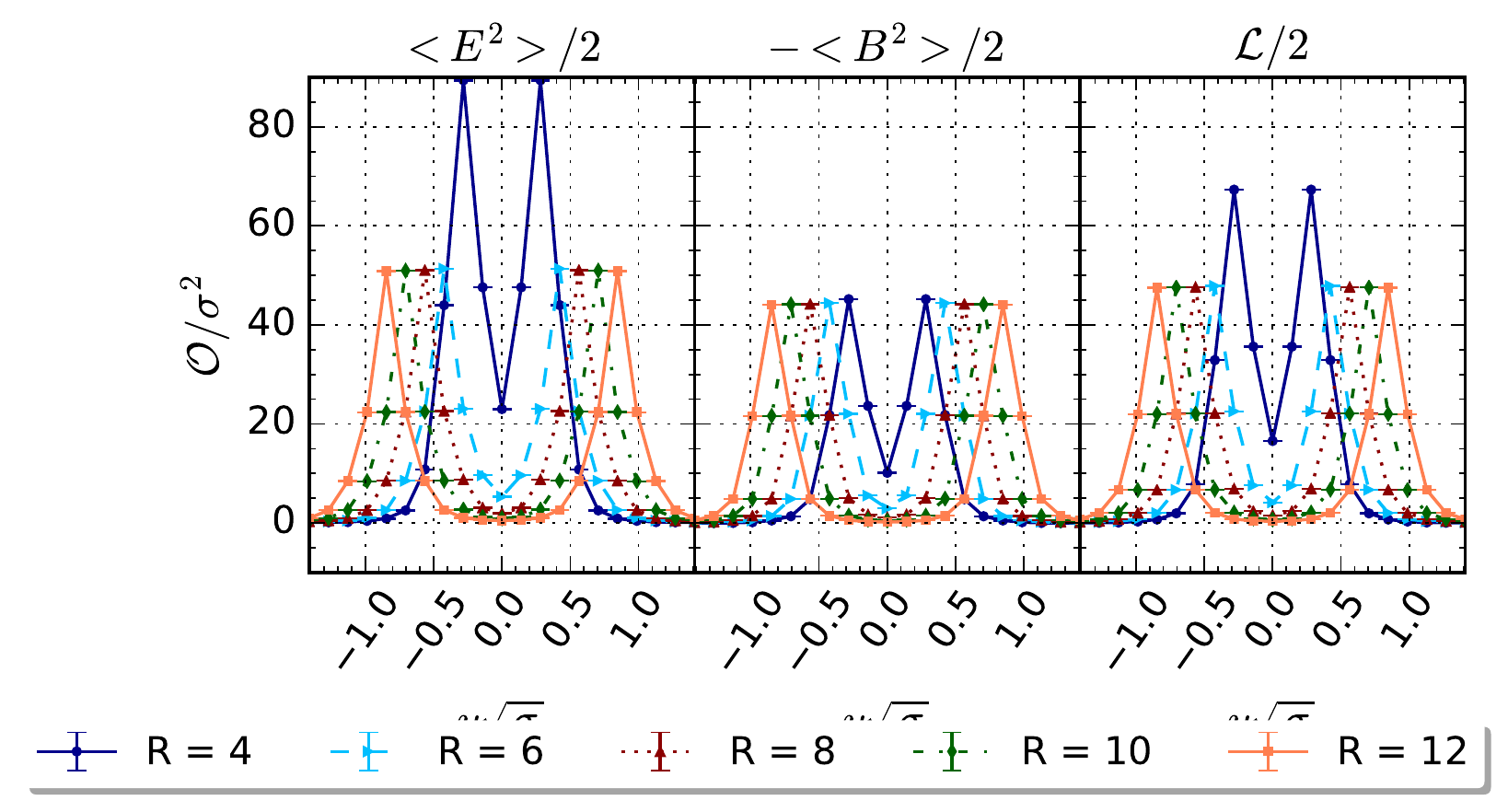}
\includegraphics[trim=0 30 10 0, clip,width=0.50\columnwidth]{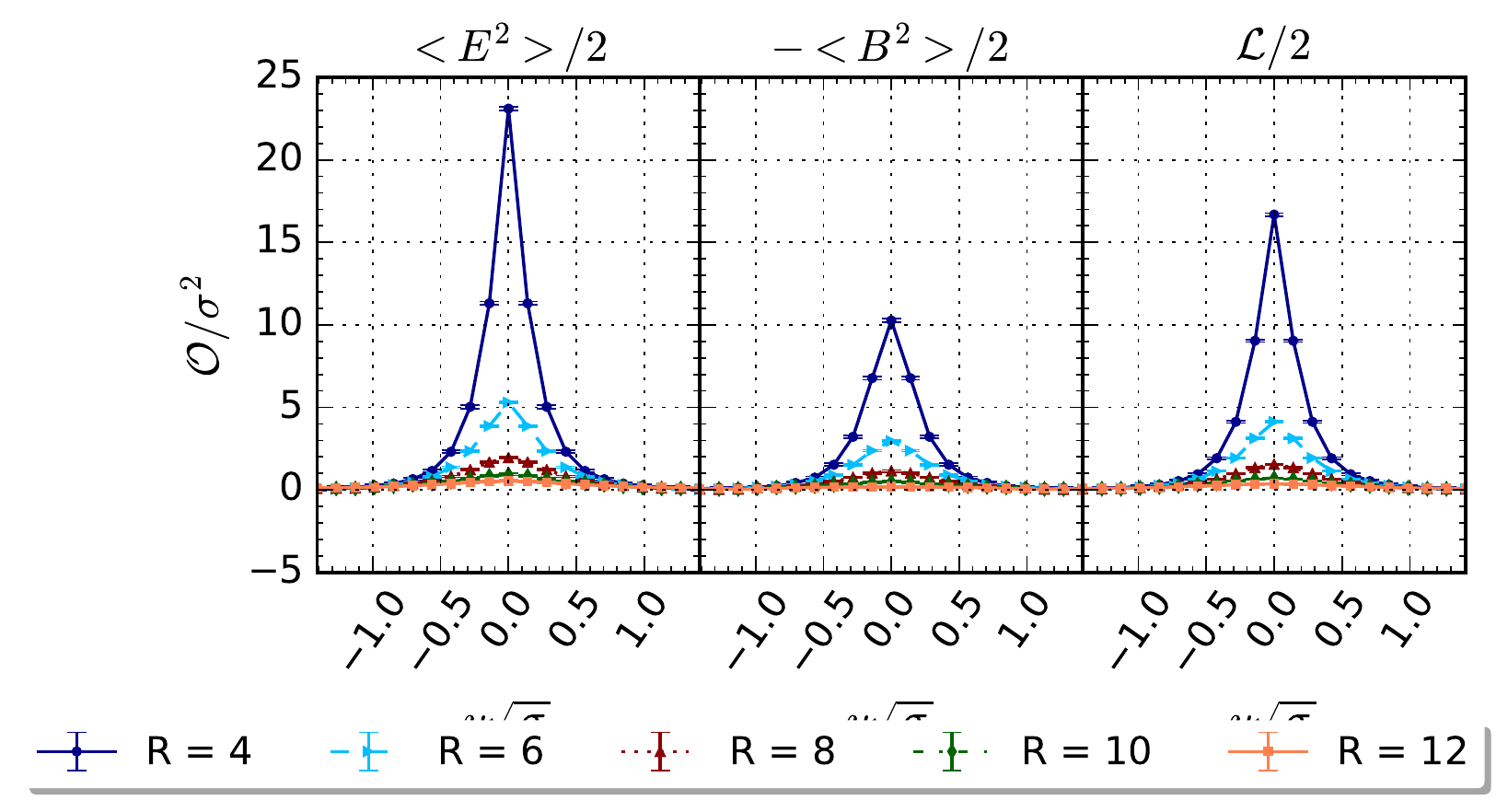}
\par\end{centering}}

	\hspace{-15pt}
\subfloat[$\beta=6.4249$, $T=1.690T_c$.\label{fig:F00_6.4249_pp_}]{
\begin{centering}
\includegraphics[trim=0 30 10 0, clip,width=0.50\columnwidth]{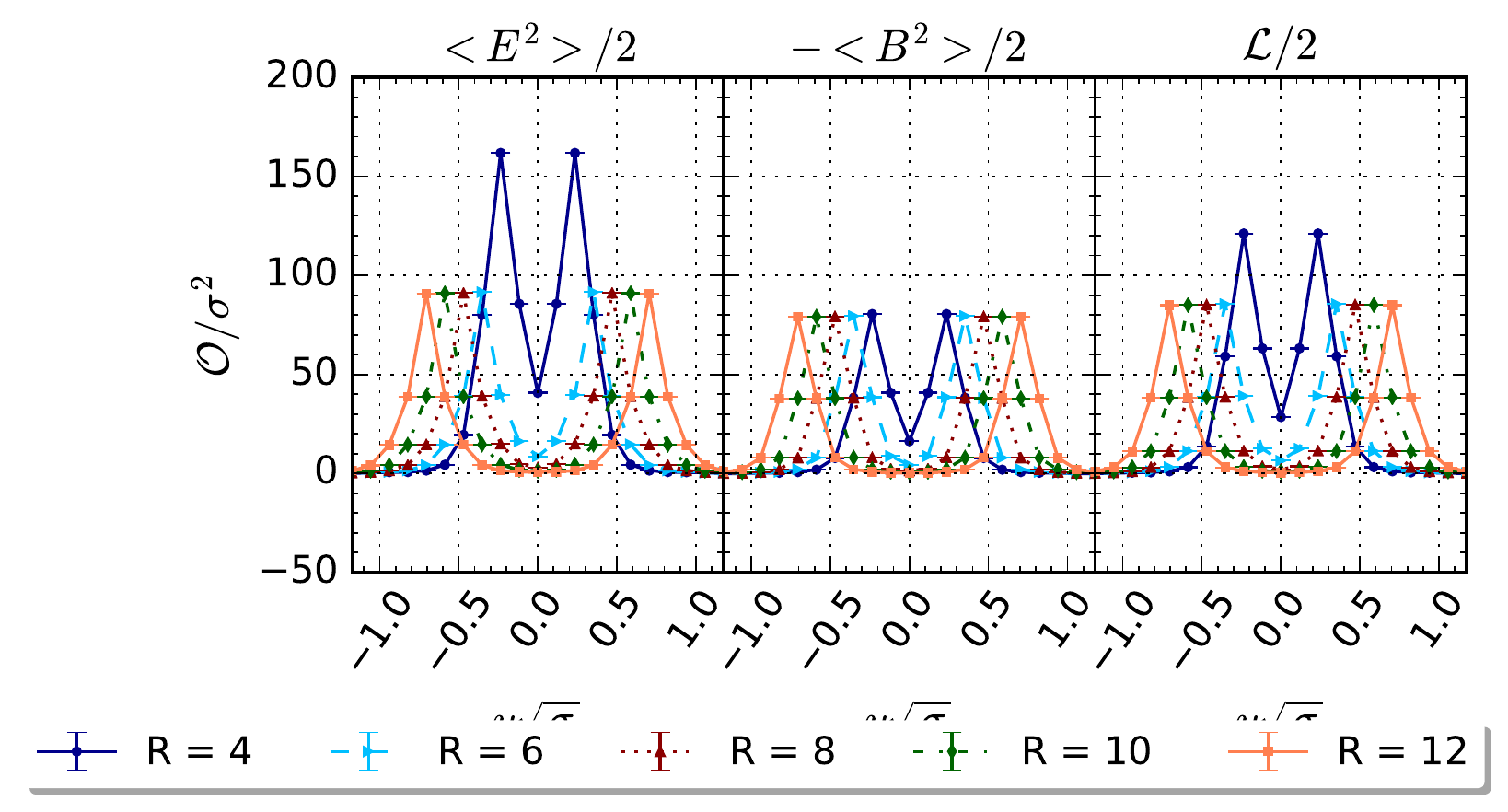}
\includegraphics[trim=0 30 10 0, clip,width=0.50\columnwidth]{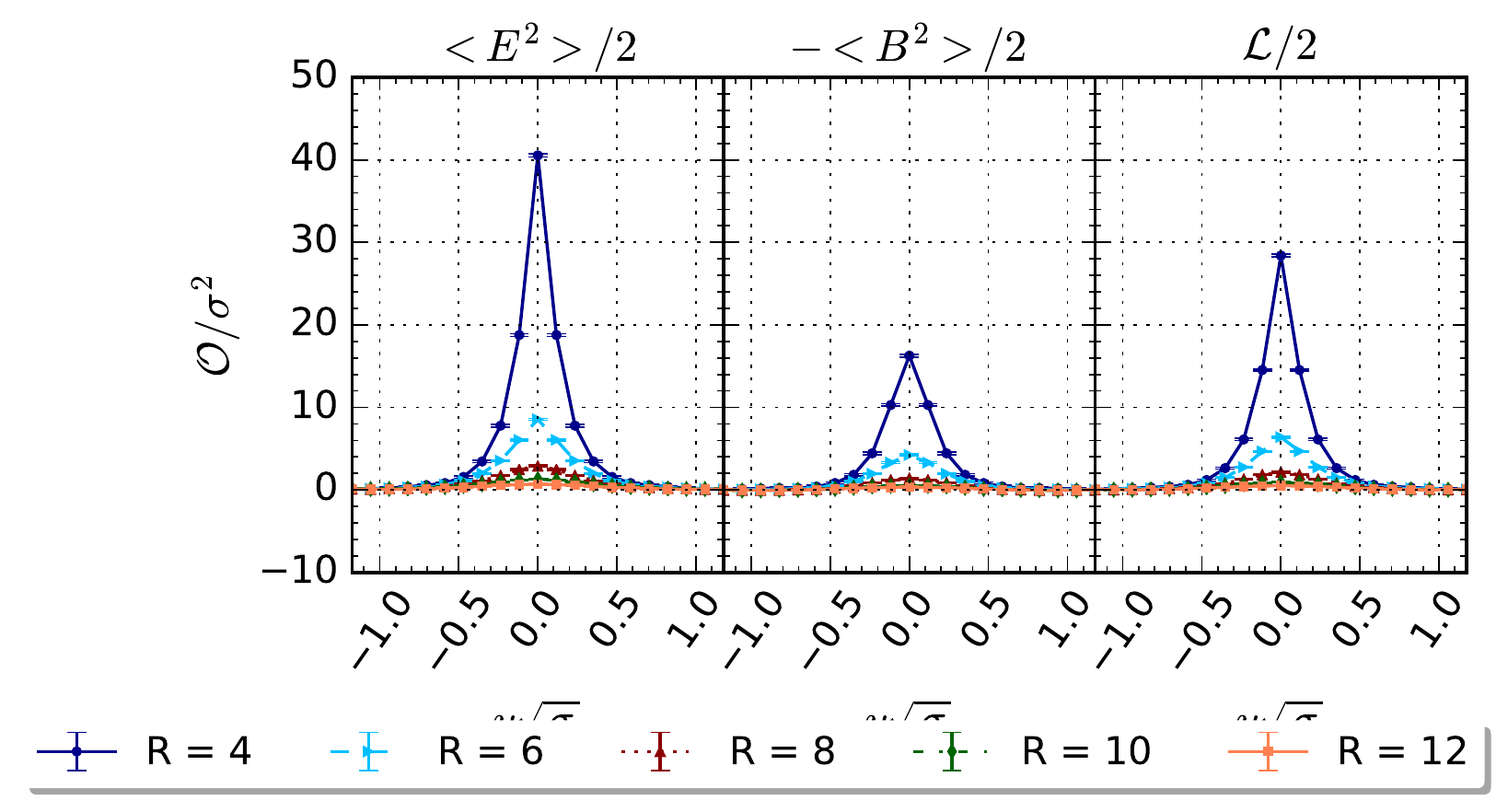}
\par\end{centering}}

\includegraphics[trim=5 0 10 260, clip,width=0.60\columnwidth]{F11_6.5_XZ_ppdagger_}
\par\end{centering}
\caption{The results for the $QQ$ system. The results in the left column correspond to the fields along the sources (plane XY) and the right column to the results in the middle of the flux tube (plane XZ). $R$ is the distance between the sources in lattice units.}
\label{fig:shapeFluxTube_QQ}
\end{figure}

The extended multihit consists in replacing each temporal link by it's thermal average with the first $N$ neighbours fixed. 
rather than just taking the thermal average of a temporal link with the first neighbours. We apply the heat-bath algorithm to all the links inside, averaging the central link,
\begin{equation}
U_4\rightarrow \bar{U}_4=\frac{\int \left[\mathcal{D}U_4\right]_\Omega U_4 \,e^{\beta \sum_{\mu,s} \Tr\left[ U_\mu(s) F^\dagger(s)\right]}}{\int \left[\mathcal{D}U_4\right]_\Omega \,e^{\beta \sum_{\mu,s} \Tr\left[ U_\mu(s) F^\dagger(s)\right]}} \ .
\end{equation}
By using $N = 2$ we are able to greatly improve the signal, when compared with the error reduction achieved with the simple multihit. Of course, this technique is more computer intensive than simple multihit, while being simpler to implement than multilevel.
The only restriction is $R > 2N$ for this technique to be valid.

Moreover, just below the phase transition, we need to make sure that we don't have contaminated configurations as already mentioned in \cite{Cardoso:2011hh}. By plotting the histogram of  Polyakov loop history for $\beta=6.055$, \cref{histpolyloop}, we are able to identify a second peak, and thus we remove all the configurations that lie on the second peak. Therefore, in \cref{tab:latticesimdata} the value with asterisk corresponds to the configurations after removing these contaminated configurations. 
%In \cref{fig:F11_6.055_pp_,fig:F11_clean_6.055_pp_}, we show the results of this effect for the  $QQ$ system below the phase transition. 

%SSSSSSSSSSSSSSSSSSSSSSSSSSSSSSSSSSSSSSSSSSSSSSSSSSSSSSSSSSSSSSSSSSSSSSSSSSSS
%SSSSSSSSSSSSSSSSSSSSSSSSSSSSSSSSSSSSSSSSSSSSSSSSSSSSSSSSSSSSSSSSSSSSSSSSSSSS
%SSSSSSSSSSSSSSSSSSSSSSSSSSSSSSSSSSSSSSSSSSSSSSSSSSSSSSSSSSSSSSSSSSSSSSSSSSSS
%SSSSSSSSSSSSSSSSSSSSSSSSSSSSSSSSSSSSSSSSSSSSSSSSSSSSSSSSSSSSSSSSSSSSSSSSSSSS
%SSSSSSSSSSSSSSSSSSSSSSSSSSSSSSSSSSSSSSSSSSSSSSSSSSSSSSSSSSSSSSSSSSSSSSSSSSSS
%SSSSSSSSSSSSSSSSSSSSSSSSSSSSSSSSSSSSSSSSSSSSSSSSSSSSSSSSSSSSSSSSSSSSSSSSSSSS
%SSSSSSSSSSSSSSSSSSSSSSSSSSSSSSSSSSSSSSSSSSSSSSSSSSSSSSSSSSSSSSSSSSSSSSSSSSSS
\section{ Results}

In this section, we present the results for different $\beta$ values suing a fixed lattice volume of $48^3\times 8$, \cref{tab:latticesimdata}. The lattice spacing was computed using the parametrization from 
Ref.  \cite{Edwards:1997xf}
 in units of the string tension at zero temperature. 
 All our computations are fully performed in NVIDIA GPUs using our CUDA codes.

The two charges, $Q$ $\bar{Q}$ or $G$, are located at $(0,-R/2,0)$ and $(0,R/2,0)$ for $R=4,6,8,10\text{ and } 12$ lattice spacing units.

In \cref{fig:shapeFluxTube_QQbar_Tless,fig:shapeFluxTube_QQbar_Tmore}, we show the results for the $Q\bar{Q}$ system. As expected the strength of the fields decrease with the temperature. Also, in the confined phase the width in the middle of the flux tube increases with the distance between the sources, while above the phase transition the width decreases with the distance.

%----------------------------------------------------------------------------------------
\begin{figure}[t!]
\hspace{-15pt}
\begin{centering}
\includegraphics[width=0.45\columnwidth]{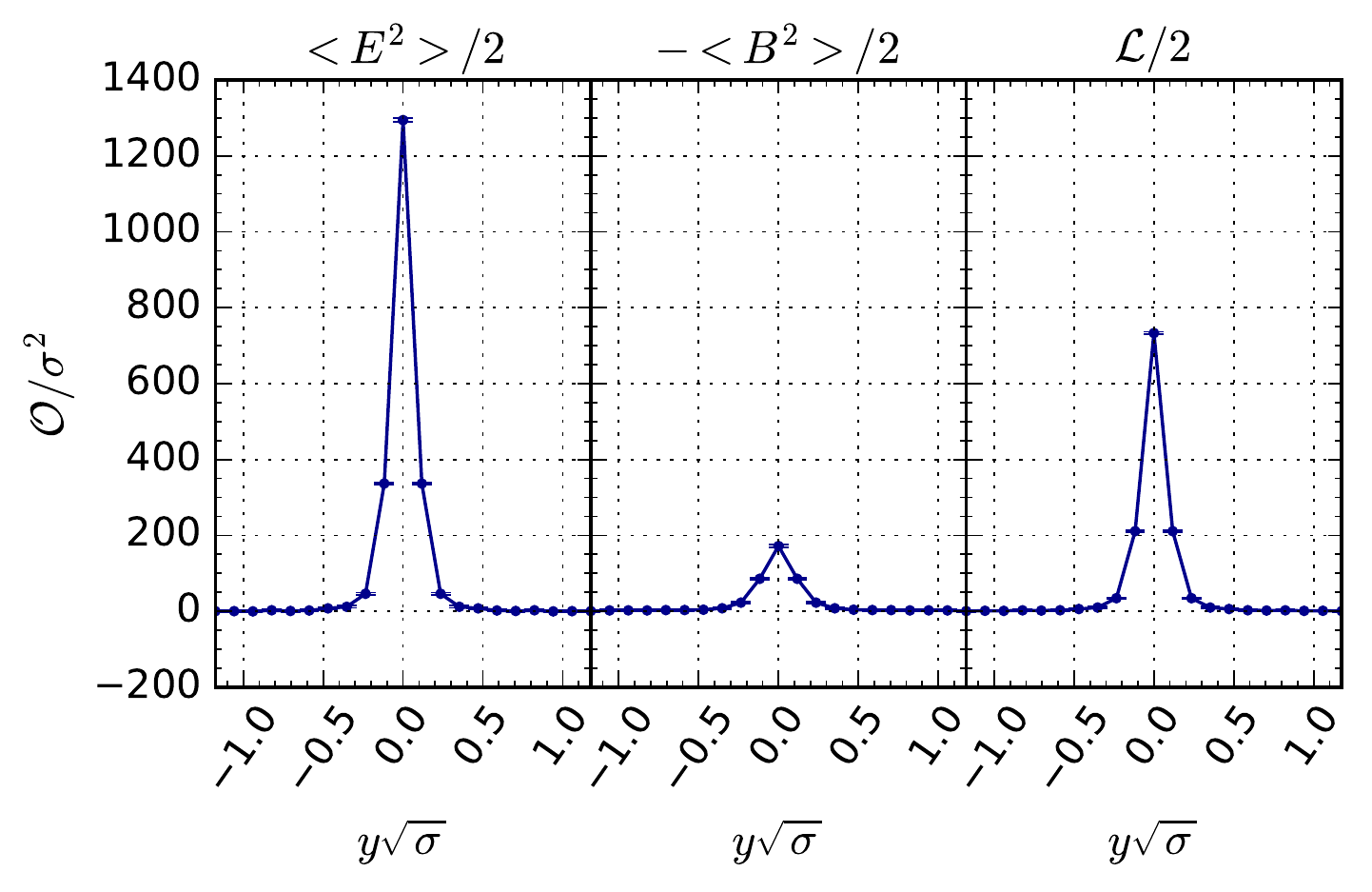}
\includegraphics[width=0.54\columnwidth]{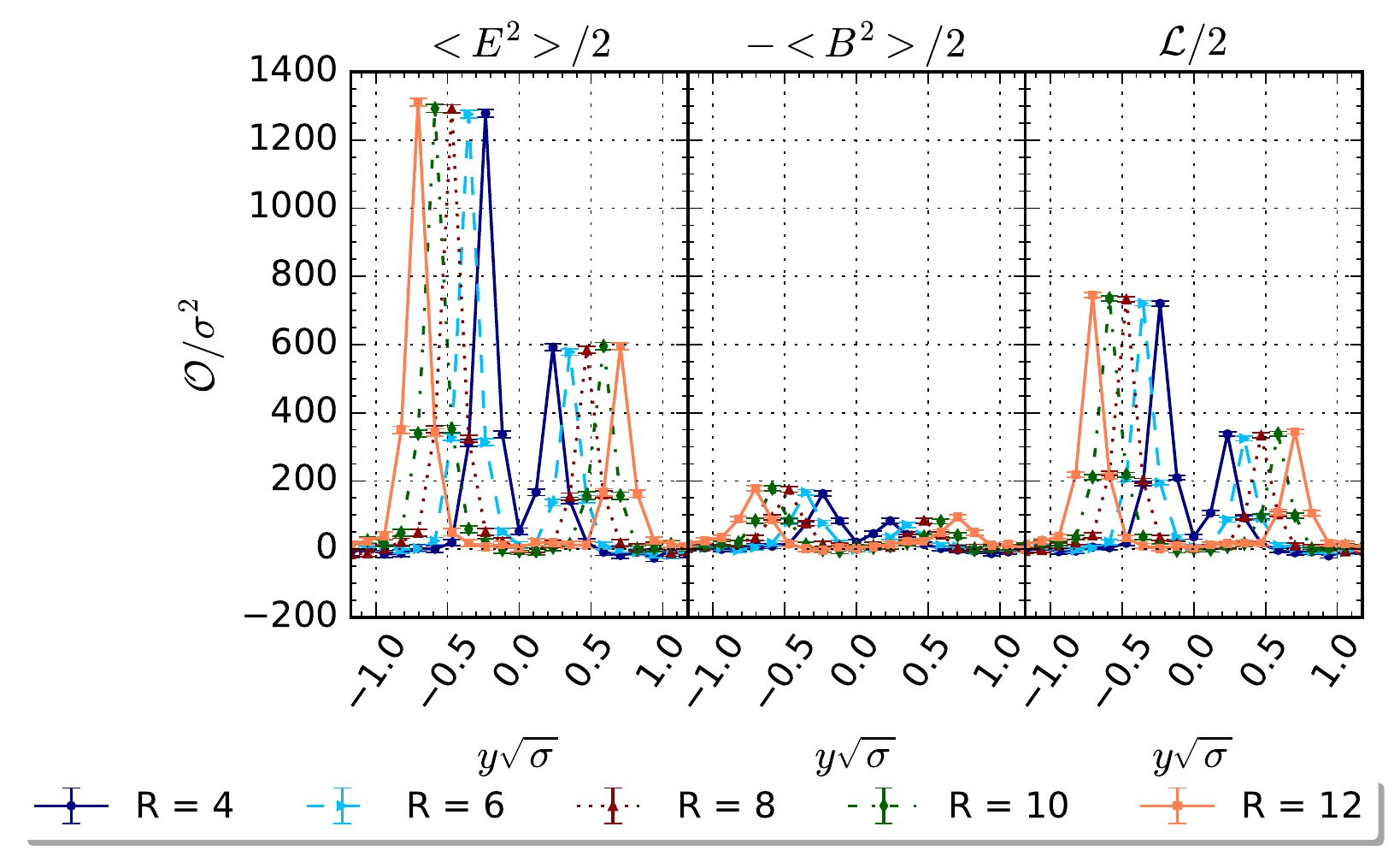}
\par\end{centering}
\caption{The results in the left correspnd to the single gluon system, and in the right to the $QG$ system, both for $\beta=6.4249$, $T=1.690T_c$.}
\label{fig:GandQGsystems}
\end{figure}

In \cref{fig:shapeFluxTube_QQ}, we show the results for the $Q Q$ system at $T > T_c$ (below $T_c$ the Polyakov loop of non colour-singlet systems vanish).  It is remarkable that these results are identical, modulo statistic errors, to the ones of the $Q\bar{Q}$ system. We interpret this as an evidence the imaginary part of the Polyakov loops is vanishing, since the Polyakov loop and anti-loop are hermitian conjugate, as in Eq. \ref{eq:polyakov_loops}, and thus only differ in the imaginary part.

Moreover, in \cref{fig:GandQGsystems} we study the effect of including a static gluon in the system with an adjoint Polyakov loop as in Eq. \ref{eq:polyakov_loops}. As in the case with quark sources only, we find evidence for no flux tube at $T > T_c$. The total square fields are similar to a simple sum of the fields produced by two charges, in qualitative agreement with the superposition property.

\section{ Conclusions}

Using CUDA codes and computations in NVIDIA GPUs only, we compute the square densities of the chromomagnetic and chromoelectric fields produced by different Polyakov loop sources, above and below the phase transition.

As the distance increase between the sources, the fields square densities decrease.
Below the deconfinement critical temperature, this decrease is moderate and is consistent with the widening of the flux tube  as already seen in studies at zero temperature \cite{Cardoso:2013lla}, moreover the field strength clearly decreases as the temperature increases as expected from the critical curve for the string tension \cite{Cardoso:2011hh}. 
Above the deconfinement critical temperature, the fields rapidly decrease to zero as the quarks are pulled apart, qualitatively consistent with screened Coulomb-like  fields.
While the width of the flux tube below the phase transition temperature increases with the separation between the quark-antiquark, above the phase transition the width seems to decrease with the separation.

As an outlook, we plan to complete the present study with a test of the cancellation of the imaginary part of the Polyakov loops at $T$ above $T_c$ and with a quantitative study of the widening of the flux tubes at $T$ below $T_c$. We also plan to produce the different Polyakov loop - Polyakov loop potentials.

\acknowledgments

Nuno Cardoso and Marco Cardoso are supported by FCT under the contracts SFRH/BPD/109443/2015 and SFRH/BPD/73140/2010 respectively.
%We would like to thank NVIDIA Corporation for the hardware donation used in this work via Academic Partnership program. ?????????
We also acknowledge the use of CPU and GPU servers of PtQCD, supported by NVIDIA, CFTP and FCT grant UID/FIS/00777/2013.

%\bibliographystyle{elsarticle-num}
%\bibliography{bib}

\end{document}